\documentclass[a4paper]{panl}
\usepackage{cite}
\usepackage{wrapfig}
\usepackage{graphicx}
\usepackage{amssymb}
\usepackage{amsfonts}
\usepackage{amsmath}
\usepackage{longtable}
\usepackage{rotating}
\usepackage{lscape}
\usepackage{epsfig}
\usepackage{multirow}

\usepackage{color} 
\usepackage{xcolor}
\usepackage{dsfont} 

\def\cmath{\color{blue}}

\newcommand{\barket}[1]{\left|#1\right\rangle} 
\def\bmat{\begin{pmatrix}}
\def\emat{\end{pmatrix}}
\newcommand{\brabar}[1]{\left\langle#1\right|} 
\def\C{\mathbb{C}} 
\newcommand{\cabs}[1]{\left|#1\right|} 
\newcommand{\CL}[1]{\mathrm{CL}\farg{#1}}	
\newcommand{\PCL}[1]{\mathrm{PCL}\farg{#1}}	
\newcommand{\KG}[1]{\mathbb{K}_{#1}} 
\newcommand{\CG}[1]{\Z_{#1}} 
\DeclareMathOperator{\diag}{diag}
\ifx\e\undefined\DeclareMathOperator{\e}{e}\fi
\newcommand{\farg}[1]{\!\left(#1\right)} 
\def\F{\mathbb{F}} 
\newcommand{\FF}[1]{\F_{#1}} 
\def\Hspace{\mathcal{H}} 
\DeclareMathOperator{\idmat}{\mathds{1}}
\DeclareMathOperator{\imu}{i}
\newcommand{\inner}[2]{\left\langle#1\mid#2\right\rangle} 
\newcommand{\Math}[1]{$\cmath{}#1$} 

\newcommand{\Mtwo}[4]{\bmat#1&#2\\#3&#4\emat} 
\newcommand{\ordset}[1]{\left[#1\right]} 
\def\runisymb{\omega} 
\def\Pbold{\mathrm{\mathbf{P}}} 
\def\Q{\mathbb{Q}} 
\newcommand{\runi}[1]{\runisymb_{#1}} 
\newcommand{\set}[1]{\left\{#1\right\}} 
\newcommand{\sympf}[2]{\left\langle#1,#2\right\rangle} 
\ifx\tr\undefined\DeclareMathOperator{\tr}{tr}\fi
\def\U{\mathsf{U}} 
\newcommand{\UG}[1]{\U\!\vect{#1}} 
\newcommand{\veco}[2]{\barket{#1}\brabar{#2}}	
\newcommand{\vect}[1]{\left(#1\right)} 
\newcommand{\Vtwo}[2]{\bmat#1\\#2\emat} 
\newcommand{\Vthree}[3]{\bmat#1\\#2\\#3\emat} 
\newcommand{\WH}[1]{\mathrm{WH}\farg{#1}}	
\newcommand{\PWH}[1]{\mathrm{PWH}\farg{#1}}	
\def\Z{\mathbb{Z}} 

\originalTeX
\begin{document}

\title{Constructive quantum mechanics based on finite groups}

\maketitle
\authors{V.V.\,Kornyak\footnote{E-mail: vkornyak@gmail.com}}
\from{LIT JINR, Dubna, Russia}

\begin{abstract}
A formulation of quantum mechanics based on replacing the general unitary group by finite groups is considered.
To solve problems arising in the context of this formulation, we use computer algebra and computational group theory  methods.
\end{abstract}
\vspace*{6pt}


\label{sec:intro}
\section*{Introduction}
A constructive version of physical theory can be built by replacing infinite sets with finite ones in the formalism.
This will not create problems with describing empirical reality, since finite sets can be arbitrarily large.
\par
The unitary group serves a dual function in standard quantum mechanics:
(1) its one-dimensional subgroups describe quantum evolutions;
(2) the group as a whole describes symmetries of quantum systems.
In the considered version of quantum mechanics, a finite group of cyclic permutations is used to describe unitary evolution instead of a continuous one-parameter group.
Taking into account the conditions ensuring quantum interference leads to a finite Weyl--Heisenberg group.
Some elements of this group, the displacement operators, generate all possible quantum evolutions in the constructive context.
The finite Clifford group, the automorphism group of the Weyl--Heisenberg group, serves as the symmetry group of quantum systems.
\par
Discarding continuous groups provides, in particular, a natural explanation for the lack of observations of quantum interference and entanglement between particles of different types.
Analysis of decompositions of a quantum system into subsystems based on the structure of a finite cyclic group shows that essentially quantum behavior is manifested only in subsystems with prime power dimensions of Hilbert spaces.
\par
Replacing continuous groups with finite ones implies a modification of the concept of quantum states.
A possible approach based on taking into account symmetries under the Clifford group and  requiring rationality of transition probabilities between states is studied.

\label{sec:CyclicEvo}
\section{Description of quantum evolution by a finite cyclic group}
In standard quantum mechanics, unitary evolution is generated by the Hamiltonian:
\Math{\displaystyle U_t=\e^{-\imu\frac{H}{\hbar}t}=\vect{\e^{-\imu\frac{H}{\hbar}}}^t=\mathrm{E}^t.}
Without loss for the description of physical reality, we can assume that time \Math{t} is an integer parameter, and the operator \Math{\mathrm{E}} is an element of the representation of the finite cyclic group \Math{\CG{N}}, where \Math{N} is a large natural number.
In \cite{Banks}, assuming that time \Math{t} is given in Planck units, estimates are given of~\Math{N\sim}~{\small$\begin{cases}
\mathrm{Exp}\farg{\mathrm{Exp}\farg{20}}&\!\!\text{for 1 cm$^3$ of matter}\\
\mathrm{Exp}\farg{\mathrm{Exp}\farg{123}}&\!\!\text{for the Universe\,.}
\end{cases}$}
\par
\textbf{Continuous vs finite group.}
The one-dimensional Lie group is the unitary group \Math{\UG{1}}, realized as the unit circle in the complex plane.
In applications, this group is approximated by the finite group \Math{\CG{N}}, which is much more complicated than \Math{\UG{1}}.
\Math{\CG{N}} decomposes into a product of cyclic groups of coprime orders:
\Math{\CG{N}\cong\CG{p_1^{\ell_1}}\times\cdots\times\CG{p_M^{\ell_M}},} 
where \Math{N=p_1^{\ell_1}\cdots{}p_M^{\ell_M}} is the product of powers of \emph{distinct} primes.
Topologically, \Math{\CG{N}} is a discrete multidimensional torus whose topology resembles the one-dimensional topology of a circle only if \Math{N} is a prime number.
The group \Math{\CG{N}} can be identified with the ring of integers modulo \Math{N} by extending the set of operations \Math{\set{+}\rightarrow\set{+,\times}}.
\par
\textbf{Galois fields and quantum mechanics.}
A group of the form \Math{\CG{p^{\ell}}} can be viewed as a substructure of the Galois field \Math{\FF{p^{\ell}}}.
The advantage of the field \Math{\FF{p^{\ell}}} over the ring \Math{\CG{p^{\ell}}} is the multiplicative invertibility of all non-zero elements.
Galois fields can be described recursively:
\par
(a) If \Math{\ell=m\cdot{n}}, then the Galois field \Math{\FF{p^{\ell}}} can be constructed as an extension of degree \Math{n} of the Galois field \Math{\FF{p^{m}}}, i.e. as an \Math{n}-dimensional vector space over \Math{\FF{p^{m}}}, the basis of which is formed by the powers of a root of an arbitrarily chosen irreducible polynomial over \Math{\FF{p^{m}}}.
Multiplication in \Math{\FF{p^{\ell}}} is defined as multiplication of elements of \Math{\FF{p^{\ell}}}, interpreted as polynomials, modulo the chosen polynomial, and addition is the usual vector addition.
A \emph{trace} relative to the extension \Math{\FF{p^{\ell}}\supset\FF{p^{m}}} is a mapping \Math{\FF{p^{\ell}}\rightarrow\FF{p^{m}}} defined for \Math{\alpha\in\FF{p^{\ell}}} by \Math{\displaystyle\tr\farg{\alpha}\equiv\tr_{\ell/m}\farg{\alpha}=\alpha+\alpha^{p^m}+\alpha^{p^{m\cdot2}}+\cdots+\alpha^{p^{m\cdot(n-1)}}\in\FF{p^m}}.
\par
(b) If \Math{m=1}, then \Math{\FF{p^{m}}} is a \emph{prime field} \Math{\FF{p}=\Z_p}.
\par
Galois fields allow us to describe \textbf{multiparticle quantum systems} of indistinguishable particles.
A quantum system with \Math{N=p^{\ell}} degrees of freedom can be represented as a collection of \Math{n} identical subsystems, each having \Math{p^m} degrees of freedom.
The variables of such a system take values in the field \Math{\FF{p^{\ell}}}, and its Hilbert space has the structure
\Math{\displaystyle\Hspace_{p^{\ell}}\cong{\Hspace_{p^m}\otimes\cdots\otimes\Hspace_{p^m}}.}
\par
\textbf{Permutation representation of \Math{\CG{N}}.}
The permutation matrix
\Math{X=\sum_{i=0}^{N-1}\veco{i+1}{i}} generates a \emph{regular} representation of the group \Math{\CG{N}} in the \Math{N}-dimensional Hilbert space.
The regular representation is also generated by the matrices \Math{X_v=X^v=\sum_{i=0}^{N-1}\veco{i+v}{i}}, \Math{\gcd\farg{v,N}=1}.
The basis \Math{\mathrm{B}_{X}=\vect{\barket{0},\ldots,\barket{N-1}}} in the Hilbert space \Math{\Hspace_N} associated with the matrix \Math{X} is called \emph{position} or \emph{ontic} ('t Hooft \cite{HooftCA}) or \emph{computational} basis.
The quantum position operator has the form
\Math{\widehat{x}=\sum_{x=0}^{N-1}x\veco{x}{x}=\diag\vect{0,\ldots,N-1}} in this basis.
The evolution of the position operator \Math{\widehat{x}_t=X_v^t\widehat{x}_0{}X_v^{-t}}, generated by \Math{X_v}, in components has the form \Math{x_t=x_0+vt \mod N}, that is, it represents ``uniform motion with velocity \Math{v}''.
\par
In the dimension \Math{N=p^{\ell}},  a generator of the cyclic group \Math{\Z_N} of the form \Math{X_{\nu}=\sum_{\gamma\in\FF{p^\ell}}\veco{\gamma+\nu}{\gamma}} is associated to each degree of freedom \Math{\nu\in\FF{p^{\ell}}}.
\par
\textbf{Decomposition into irreducible components.}
The regular representation of a finite group contains all irreducible representations.
For \Math{\CG{N}} these representations are one-dimensional and are generated by powers of the element \Math{\runi{}=\e^{{{2\pi\imu}/{N}}}}.
The decomposition of \Math{X} into a direct sum of generators of irreducible representations has the form \Math{{}Z=FXF^{-1}=\diag\vect{1,\runi{},\ldots,\runi{}^{N-1}}}, where \Math{F} is the Fourier matrix: \Math{(F)_{ij}=\frac{1}{\sqrt{N}}\runi{}^{ij},~i,j=0,\ldots,N-1}.
Associated with the matrix \Math{Z} is the \emph{momentum basis} \Math{\mathrm{B}_Z=\vect{\barket{\widetilde{0}}, \barket{\widetilde{1}},\ldots,\barket{\widetilde{N-1}}}}, which is the Fourier transform of the basis \Math{\mathrm{B}_{X}}.
\par
If \Math{N=p^{\ell}}, then for \Math{\mu\in\FF{p^{\ell}}} we have
\Math{Z_{\mu}=\sum_{\gamma\in\F{p^\ell}}\exp\farg{\frac{2\pi\imu}{p}\tr\vect{\mu\gamma}}\veco{\gamma}{\gamma}}.
\par
A cyclic group does not have \textbf{projective representations needed to describe quantum interferences}.
However, \Math{X} and its diagonal form \Math{Z} together generate a projective representation of the group \Math{\CG{N}\times\CG{N}} in \Math{\Hspace_N}.
A direct calculation leads to the commutation relation \Math{ZX=\runi{}XZ}, which Weyl obtained by noting that the Heisenberg commutation relation for the position and momentum operators \Math{\ordset{\hat{x},\hat{p}}=\imu\hbar\idmat}, and hence the standard quantum theory as a whole, can only be realized in an infinite-dimensional Hilbert space.
Weyl's analysis of quantum behavior in \Math{N}-dimensional space \cite{WeylEn} necessarily leads to the matrices \Math{X} and \Math{Z}.
\par
The bases \Math{\mathrm{B}_X} and \Math{\mathrm{B}_Z} are \textbf{mutually unbiased}, i.e. Born's probabilities of transitions between elements of different bases are the same for any pairs of elements: \Math{\cabs{\inner{\widetilde{\ell}}{k}}^2=1/N,~~\ell,k=0,\ldots,N-1}.
This means that measuring a quantum state that is a vector of one of the bases in another basis will not provide any information: the measurement results will be scattered with equal probability across all \Math{N} possibilities.
In fact, the concept of mutually unbiased bases, first clearly formulated by Schwinger \cite{Schwinger1960}, is a mathematical formalization of Bohr's complementarity principle.
The ideas of Weyl and Schwinger are actively developed in various fields: the foundations of quantum theory, quantum information science, signal processing theory, etc.
Ignoring for brevity the specifics of ``Galois dimensions'', we describe the main elements of the formalism for the case of the ring \Math{\Z_N}.
\par
The element \Math{\tau=-\e^{\pi\imu/N}} generates \Math{\KG{N}} if \Math{N=2k+1} and \Math{\KG{2N}} if \Math{N=2k}, where \Math{\KG{n}} denotes the group of \Math{n}th roots of unity. 
The elements \Math{\tau}, \Math{X} and \Math{Z} generate the \textbf{Weyl--Heisenberg group} \Math{\WH{N}=\langle\tau, X, Z\rangle}.
The order of \Math{\WH{N}} is \Math{N^3} or \Math{2N^3} depending on the parity of \Math{N}.
Quantum evolutions are generated by the \emph{displacement operators} \Math{\mathrm{D}_{\mathbf{p}}=\tau^{p_1p_2}X^{p_1}Z^{p_2}}, \Math{\mathbf{p}=\Vtwo{p_1}{p_2}\in\Z^2}, which form the \textbf{projective Weyl--Heisenberg group} \Math{\PWH{N}} of order \Math{N^2}.
Composition of displacement operators \Math{\displaystyle\mathrm{D}_{\mathbf{p}}\mathrm{D}_{\mathbf{q}}=\tau^{\sympf{\mathbf{p}}{\mathbf{q}}}\mathrm{D}_{\mathbf{p}+\mathbf{q}}} contains the symplectic form \Math{\sympf{\mathbf{p}}{\mathbf{q}}=p_2q_1-p_1q_2}.
The symmetry group of this form, the symplectic group \Math{\mathrm{Sp}\farg{2,\Z_{N}}}, is the outer automorphism group of \Math{\WH{N}}.
\par
Combining inner and outer automorphisms, we arrive at a semidirect product called the \textbf{Clifford group} \Math{\CL{N}
\cong\WH{N}\rtimes\mathrm{Sp}\farg{2,\Z_{N}}}.
Traditionally, the Clifford group is defined as the normalizer of the Weyl--Heisenberg group in the unitary group \Math{\UG{N}}.
The need for \Math{\UG{N}}, which remains as a relic of the continuous theory, follows neither from the description of quantum evolution by finite cyclic groups, nor from Weyl's considerations.
We will consider the Clifford group exclusively as the symmetry group of the Weyl--Heisenberg group without resorting to a reference to the continuous group \Math{\UG{N}}.
The Clifford group is generated by the matrices \Math{X}, \Math{F} and \Math{S=\diag\farg{\tau^{i\vect{i+N}}, i=0,\ldots,N-1}}: \Math{\CL{N}=\langle X, F, S\rangle}.
\par
The \textbf{projective Clifford group} -- the quotient group of \Math{\CL{N}} by its center -- is generated by the same elements, \Math{\PCL{N}=\langle X, F, S\rangle}, but matrices that differ only by a phase factor are equivalent.
\section{Decomposition of a \Math{N}-dimensional quantum system into subsystems}\label{Qdeco-sec}
The Hilbert space of the \emph{global} system decomposes into a tensor product of \emph{local} spaces \Math{\Hspace_N=\Hspace_{n_1}\otimes\ldots\otimes\Hspace_{n_M}} if the global dimension decomposes into a product of coprime numbers, for example, \Math{N=n_1\cdot\ldots\cdot{}n_M}, where \Math{n_i=p_i^{\ell_i}} are powers of distinct primes.
The equivalence class of this decomposition with respect to arbitrariness in the choice of coordinates in Hilbert spaces can be symbolically described as \Math{G\farg{N}\Hspace_N=G\farg{n_1}\Hspace_{n_1}\otimes\ldots\otimes{}G\farg{n_M}\Hspace_{n_M}}, where \Math{G\farg{n}} is the symmetry group of an \Math{n}-dimensional space.
Using the properties of the \Math{\otimes} operation, the description can be simplified to \Math{G\farg{N}\Hspace_N=\Hspace_{n_1}\otimes\ldots\otimes\Hspace_{n_M}}.
That is, the decompositions lying on the orbit of the global group \Math{G\farg{N}} are equivalent and completely determined by the dimension decomposition.
\par
The assumption \Math{G\farg{N}=\UG{N}} may lead to artifacts because the continuous group \Math{\UG{N}} freely ``mixes'' states between different components of the tensor product, which would result in entanglement between fundamental, i.e. non-composite, particles of different types that is not observed in nature.
\par
The assumption \Math{G\farg{N}=\CL{N}} does not cause problems of this kind, since there are no transformations in the Clifford group of the global system that mix states between local Hilbert spaces of coprime dimensions.
Mathematically, this is expressed by the fact, proved using the Chinese remainder theorem, that the global Clifford group decomposes into a direct product of local ones: \Math{\CL{N}=\CL{n_1}\times\cdots\times\CL{n_M}}.
\par
The Chinese theorem also implies the relationship between the energy levels of the system and subsystems \Math{E_{k/N}=E_{k_1/n_1}+\ldots+E_{k_M/n_M}}, where \Math{E_{\nu}=h\nu}.
This means that the energy of the global system is equal to the sum of the energies of the components, and there are no interaction energies.
\par
Thus, subsystems of coprime dimensions can be studied separately and independently of each other, since there are no quantum entanglement or energy interaction between them.
\par
In systems of prime dimensions, \Math{N=p}, due to the absence of subsystems, quantum entanglement is impossible.
Therefore, the main interest for study is dimensions of the form \Math{N=p^{\ell},~\ell>1}.
\section{Constructive quantum states}\label{QSTATES-sec}
In continuous quantum mechanics, the set of pure states in \Math{N}-dimensional Hilbert space is the complex projective space \Math{\Pbold\farg{\Hspace_N}=\C\Pbold^{N-1}}, which is a homogeneous space of the unitary group \Math{\UG{N}}.
This means that \Math{\C\Pbold^{N-1}} is the orbit of an arbitrary unit vector: \Math{\C\Pbold^{N-1}\cong\mathrm{Orb}_{_{\UG{N}}}\farg{\barket{0}}=\UG{N}\barket{0}}.
\par
Replacing \Math{\UG{N}} with \Math{\CL{N}} as the symmetry group, we assume that the constructive set of pure quantum states \Math{\mathrm{CQS}\farg{N}} consists of elements of the form \Math{\barket{a}=\sum_{i=0}^{N-1}\varphi_{i}\alpha_{i}\barket{i},} where \Math{\alpha_{i}\in\mathbb{R}}, \Math{\sum_{i=0}^{N-1}\alpha_{i}^2=1}, \Math{\varphi_{i}\in\mathrm{Z}\farg{\CL{N}}}, i.e. the phase factors belong to the center of \Math{\CL{N}}.
The set \Math{\mathrm{CQS}\farg{N}} must 1) be \Math{\CL{N}}-invariant; 2) contain ontic vectors; 3) consist only of elements with rational Born probabilities of transitions between each other.
Formally: 
1) \Math{\barket{a}\in\mathrm{CQS}\farg{N}\implies\mathrm{Orb}_{_{\CL{N}}}\farg{\barket{a}}\subseteq\mathrm{CQS}\farg{N}};
2) \Math{\barket{0}\in\mathrm{CQS}\farg{N}};\\
3) \Math{\barket{a},\barket{b}\in\mathrm{CQS}\farg{N}\implies\cabs{\inner{a}{b}}^2\in\Q}.
\par
To study the properties of quantum states that meet these requirements, we implemented a procedure for their sequential construction.
The initial states are constructed as the orbit of the vector \Math{\barket{0}}.
Then the process of adding new states is repeated, which boils down to obtaining quantum superpositions of existing states and selecting those superpositions whose transition probabilities are rational.
\par
\textbf{Examples of computer experiments in dimensions 2 and 3:}\\
The generators, centers and sizes of Clifford groups in these dimensions are given in the table (\Math{\runi{}=\exp\farg{2\pi\imu/3}})\\
\centerline{\begin{tabular}{c|ccc|c|c}
    \Math{N}&\Math{X}&\Math{F}&\Math{S}& \Math{\mathrm{center}}
       & \Math{\mathrm{ord}}
     \\\hline&&&&\\[-12pt] 
    \Math{2}&\Math{\Mtwo{0}{1}{1}{0}}&\Math{\!\!\!\frac{1}{\sqrt{2}}\Mtwo{1}{1}{1}{-1}}&\Math{\!\!\!\!\!\!\Mtwo{1}{0}{0}{\imu}}&\Math{\KG{8}
    }&\Math{192}\\&&&&\\[-12pt]
    \Math{3}&\Math{\hspace{-2pt}\bmat
    0&0&1\\
    1&0&0\\
    0&1&0
    \emat}&\Math{\hspace{-2pt}\frac{1}{\sqrt{3}}\!\bmat
    1&1&1\\
    1&\runi{}&\runi{}^2\\
    1&\runi{}^2&\runi{}
    \emat}&\Math{\hspace{-9pt}\bmat
    1&0&0\\
    0&\runi{}^2&0\\
    0&0&\runi{}^2
    \emat}&\Math{\KG{12}
    }&\Math{2592}
  \end{tabular}
  }
$\mathbf{N=2.}$ The results of the calculations can be demonstrated visually, since the pure states belong to the complex projective line \Math{\C\Pbold^1}, which can be represented as the Riemann (Bloch) sphere.
The projective Clifford group \Math{\PCL{2}=\CL{2}/\KG{8}} has order \Math{24}.
The possible sizes of orbits are divisors of this number.
The orbit \Math{\mathrm{Orb}_{_{\CL{2}}}\farg{\barket{0}}} consists of six vectors \Math{\barket{0}}, \Math{\barket{1}}; \Math{\displaystyle\frac{\barket{0}+\barket{1}}{\sqrt{2}}}, \Math{\displaystyle\frac{\barket{0}-\barket{1}}{\sqrt{2}}}; \Math{\displaystyle\frac{\barket{0}+\imu\barket{1}}{\sqrt{2}}}, \Math{\displaystyle\frac{\barket{0}-\imu\barket{1}}{\sqrt{2}}}, orthogonal pairs of which form a complete set%
\footnote{In dimension \Math{N=p^{\ell}} there always exists a complete set of \Math{N+1} mutually unbiased bases.
Measurements relative to vectors from these bases are sufficient for the complete reconstruction of any quantum state, pure or mixed.}
of mutually unbiased bases.
In Fig. \ref{fig01} (a) these six vectors form the vertices of an octahedron, and the spatial diagonals of the octahedron represent three mutually unbiased bases.
\begin{figure}[h]
    \begin{center}
    \includegraphics[width=0.32\textwidth]{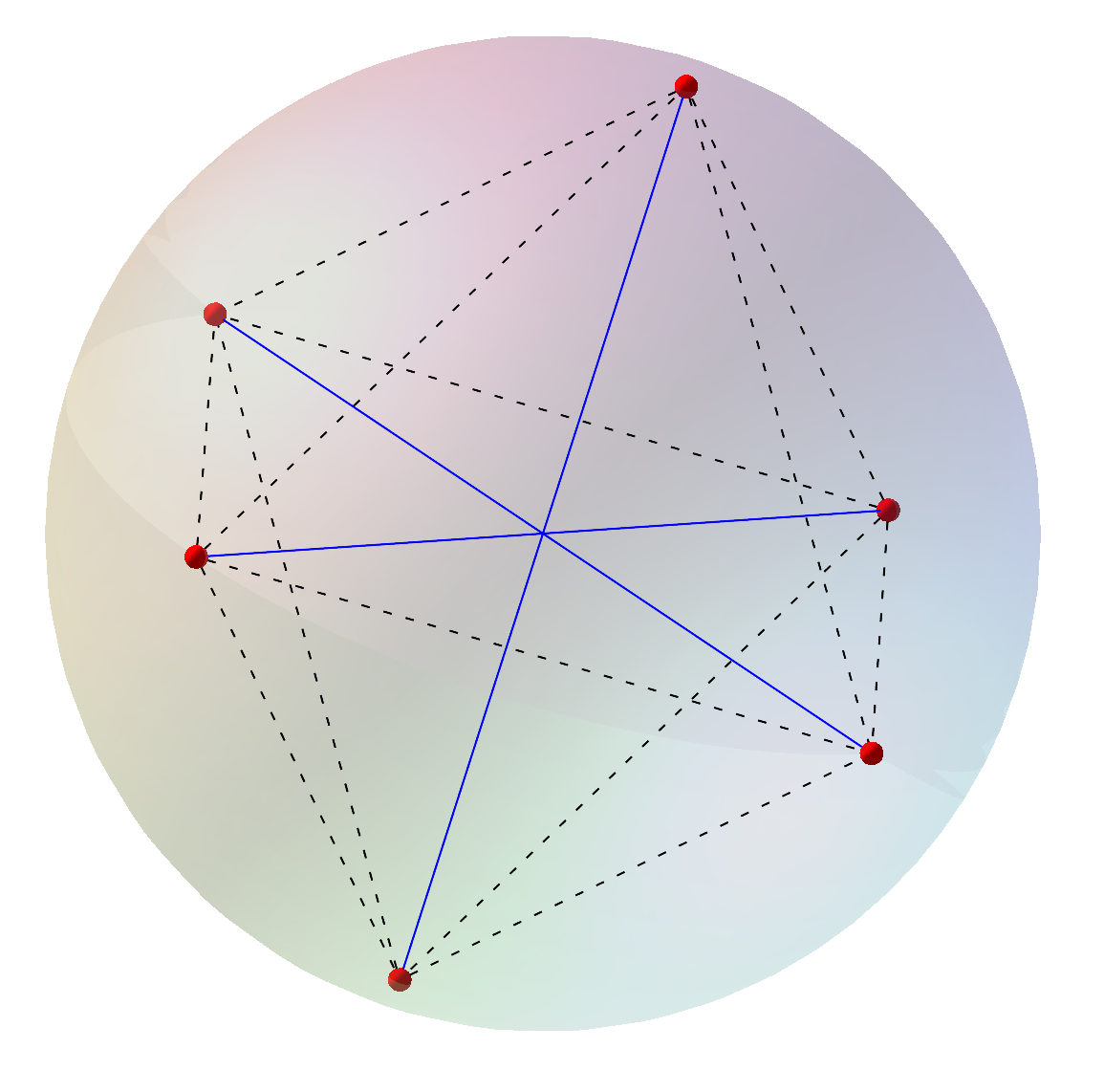}~~
    \includegraphics[width=0.31\textwidth]{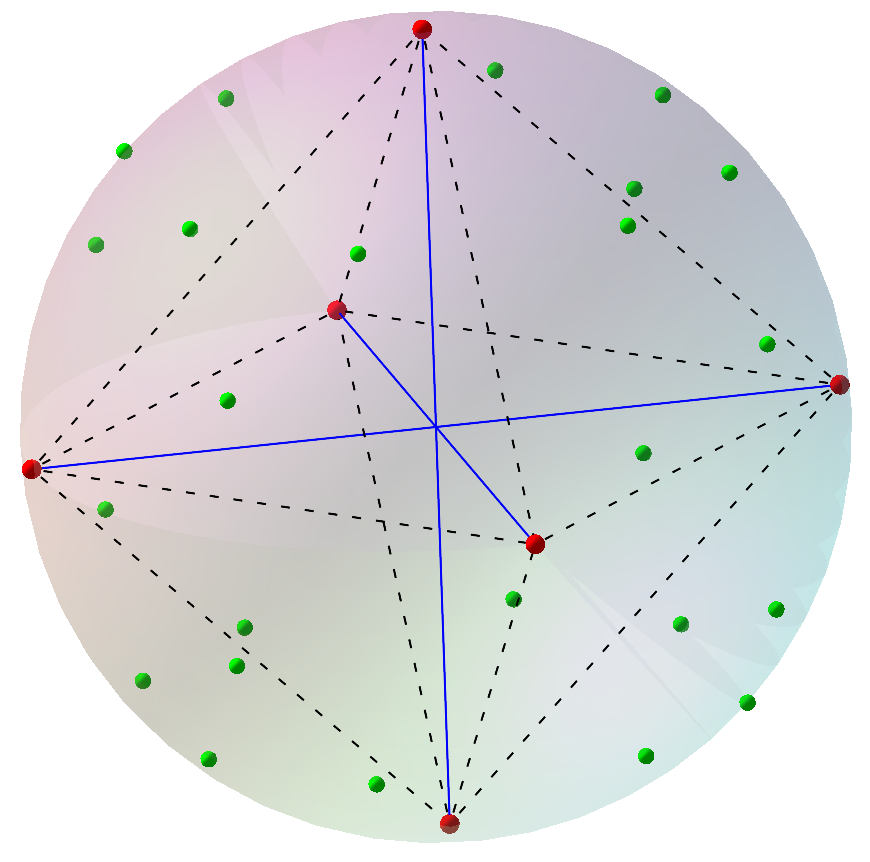}~~
    \includegraphics[width=0.3\textwidth]{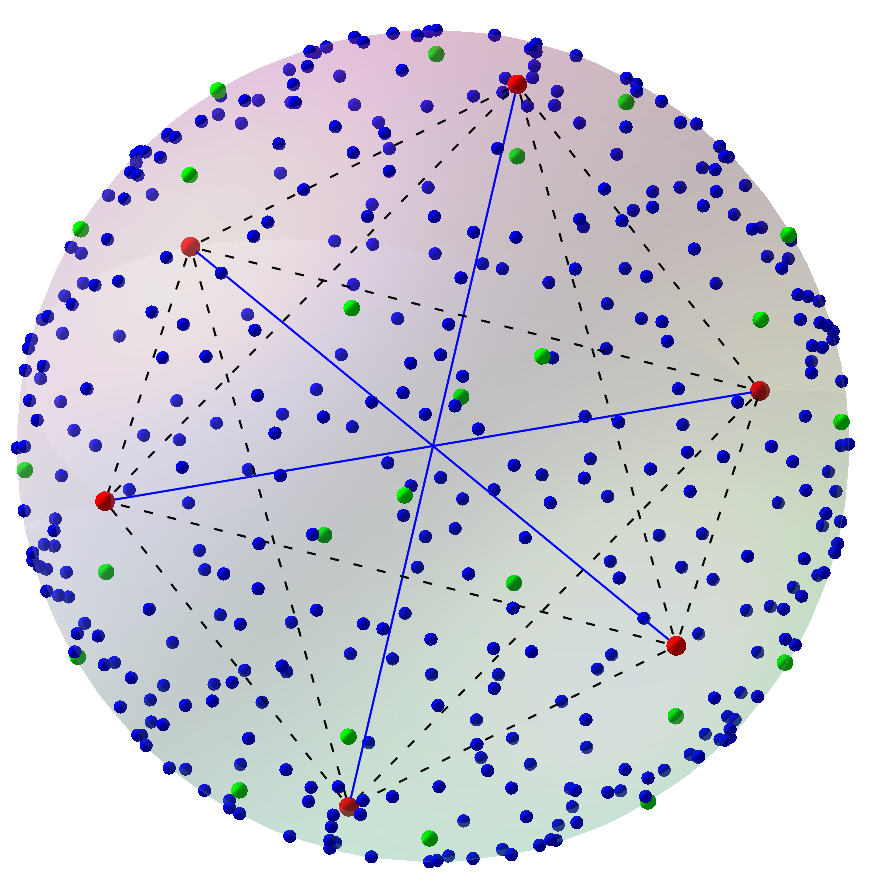}~~\\
    (a)\hspace{0.32\textwidth}(b)\hspace{0.3\textwidth}(c)
    \vspace{-3mm}
    \caption{Generating constructive quantum states}
    \end{center}
    \labelf{fig01}
    \vspace{-5mm}
  \end{figure}
The pairwise interferences of these six vectors form \Math{48} vectors, half of which are rejected due to ``incommensurability'' with the already constructed elements of \Math{\mathrm{CQS}\farg{2}}, i.e. the transition probabilities turn out to be irrational.
The remaining states form a Clifford group orbit of size \Math{24}.  
Fig. \ref{fig01} (b) shows the set of states supplemented by these vectors. 
The next step results in adding states that form \Math{16} orbits of size \Math{24} (Fig. \ref{fig01} (c)).
\par
$\mathbf{N=3.}$
The order of the projective group \Math{\PCL{3}=\CL{3}/\KG{12}} is \Math{216}.
\Math{\mathrm{Orb}_{_{\CL{3}}}\farg{\barket{0}}} consists of \Math{12} vectors \Math{\barket{0},\barket{1},\barket{2}};
\Math{\frac{1}{\sqrt{3}}\Vthree{1}{1}{1}},
\Math{\frac{1}{\sqrt{3}}\Vthree{1}{\runi{}}{\runi{}^2}},
\Math{\frac{1}{\sqrt{3}}\Vthree{1}{\runi{}^2}{\runi{}}};
\Math{\frac{1}{\sqrt{3}}\Vthree{1}{\runi{}^2}{\runi{}^2}}, 
\Math{\frac{1}{\sqrt{3}}\Vthree{1}{1}{\runi{}}},
\Math{\frac{1}{\sqrt{3}}\Vthree{1}{\runi{}}{1}};
\Math{\frac{1}{\sqrt{3}}\Vthree{1}{\runi{}}{\runi{}}},
\Math{\frac{1}{\sqrt{3}}\Vthree{1}{1}{\runi{}^2}},
\Math{\frac{1}{\sqrt{3}}\Vthree{1}{\runi{}^2}{1}}.
Consecutive triplets of these vectors form a complete set of four mutually unbiased bases.
Pairwise interferences of the vectors of this orbit generate \Math{153} vectors that form the union of three Clifford group orbits of sizes \Math{9=3^2}, \Math{36=2^23^2}, and \Math{108=2^23^3}.
\section{Conclusion}
The continuously infinite unitary group, due to its non-constructiveness, is a potential source of artifacts when used in the formalism of quantum mechanics.
We have shown that quantum behavior can be described using only finite groups.
More specifically, quantum evolutions are described by cyclic subgroups of the Weyl--Heisenberg group, and the symmetry group of quantum systems is the Clifford group.
Restricting unitary symmetries to these groups has empirically significant consequences.
In particular, the absence of quantum entanglement and interference between elementary particles of different types finds a natural explanation.
\par
Decomposition of a quantum system into subsystems is determined by the decomposition of the dimension of its Hilbert space into a product of integers.
In a prime dimension, decomposition into subsystems is impossible.
When the dimension is a product of coprime numbers, the quantum behavior of the subsystems can be studied separately, since there are no energy interactions and quantum correlations between them.
Quantum behavior manifests itself entirely in prime power dimensions.
This is the case of multiparticle quantum systems consisting of entangled indistinguishable particles.
\par
The rejection of the continuous unitary group implies a modification of the concept of a quantum state: the projective Hilbert space should be replaced by some combinatorial construction.
We assume that the set of states must satisfy the following requirements:
(a) the phase factors in the states must be elements of the center of the Clifford group;
(b) the set of states must be invariant under the Clifford group and (c) contain ontic vectors;
(d) the Born probabilities of transitions between elements of the set must be rational in accordance with the frequency concept of probability.
Some results of computer experiments with this set of requirements are presented.

\end{document}